\documentclass[%
 reprint,
superscriptaddress,
nofootinbib,
 amsmath,amssymb,amsfonts,
aps,
]{revtex4-1}

\usepackage{graphicx}
\usepackage{dcolumn}
\usepackage{bm, bbold}
\usepackage{comment, color}
\usepackage{natbib}
\usepackage{tabularx}

\begin{document}


\def\bea{\begin{eqnarray}}
\def\eea{\end{eqnarray}}
\def\beq{\begin{equation}}
\def\eeq{\end{equation}}
\def\f{\frac}
\def\k{\kappa}
 
\def\ve{\varepsilon}
\def\be{\beta}
\def\D{\Delta}
\def\h{\theta}
\def\t{\tau}
\def\a{\alpha}
\def\e{\D_{\theta}}
\def\cDa{{\cal D}[X]}
\def\cD{{\cal D}[x]}
\def\cL{{\cal L}}
\def\cLo{{\cal L}_0}
\def\cLa{{\cal L}_1}
\def\rv{{\bf r}}
\def\tv{\hat t}
\def\on{{\omega_{\rm a}}}
\def\od{{\omega_{\rm d}}}
\def\off{{\omega_{\rm off}}}
\def\fv{{\bf{f}}}
\def\fm{\bf{f}_m}
\def\zh{\hat{z}}
\def\yh{\hat{y}}
\def\xh{\hat{x}}
\def\km{k_{m}}
\def\nh{\hat{n}}

\def\Re{{\rm Re}}
\def\sj{\sum_{j=1}^2}
\def\rk{\rho^{ (k) }}
\def\rek{\rho^{ (1) }}
\def\cek{C^{ (1) }}
\def\rz{\rho^{ (0) }}
\def\rt{\rho^{ (2) }}
\def\rtb{\bar \rho^{ (2) }}
\def\trk{\tilde\rho^{ (k) }}
\def\trek{\tilde\rho^{ (1) }}
\def\trz{\tilde\rho^{ (0) }}
\def\trt{\tilde\rho^{ (2) }}
\def\r{\rho}
\def\tD{\tilde {D}}

\def\kb{k_B}
\def\bF{\bar{\cal F}}
\def\F{{\cal F}}
\def\la{\langle}
\def\ra{\rangle}
\def\nn{\nonumber}
\def\up{\uparrow}
\def\dn{\downarrow}
\def\S{\Sigma}
\def\s{\sigma}
\def\dg{\dagger}
\def\d{\delta}
\def\p{\partial}
\def\l{\lambda}
\def\L{\Lambda}
\def\G{\Gamma}
\def\o{\Omega}
\def\w{\omega}
\def\g{\gamma}
\def\E{{\mathcal E}}

\def\O{\Omega}

\def\vv{ {\bf v}}
\def\jv{ {\bf j}}
\def\jr{ {\bf j}_r}
\def\jd{ {\bf j}_d}
\def\jdd{ { j}_d}
\def\noi{\noindent}
\def\a{\alpha}
\def\d{\delta}
\def\p{\partial} 

\def\la{\langle}
\def\ra{\rangle}
\def\e{\epsilon}
\def\n{D_\h}
\def\g{\gamma}
\def\hf{\frac{1}{2}}
\def\rcurs{r_{ij}}

\def\ev{ {\hat {\bf e}}}
\def\bv{ {\bf b}}
\def\uv{ {\bf u}}
\def\rv{ {\bf r}}
\def\r0{ {\bf 0}}
\def\cf{{\mathcal F}}
\def\e{D_\h}




\title{How reciprocity impacts ordering and phase separation in active nematics? }

\author{Arpan Sinha}
\email{arpan.s@iopb.res.in}
\affiliation{Institute of Physics, Sachivalaya Marg, Bhubaneswar 751005, India}
\affiliation{Homi Bhaba National Institute, Training School Complex, Anushakti Nagar, Mumbai 400094, India}

\author{Debasish Chaudhuri}
\email{debc@iopb.res.in}
\affiliation{Institute of Physics, Sachivalaya Marg,  Bhubaneswar 751005, India}
\affiliation{Homi Bhaba National Institute, Training School Complex, Anushakti Nagar, Mumbai 400094, India}

\begin{abstract}
Active nematics undergo spontaneous symmetry breaking and show phase separation instability. Within the prevailing notion that macroscopic properties depend only on symmetries and conservation laws, different microscopic models are used out of convenience. Here, we test this notion carefully by analyzing three different microscopic models of apolar active nematics. They share the same symmetry but differ in implementing reciprocal or non-reciprocal interactions, including a Vicsek-like implementation. We show how such subtle differences in microscopic realization determine if the ordering transition is continuous or first order. Despite the difference in the type of phase transition, all three models exhibit fluctuation-dominated phase separation and quasi-long-range order in the nematic phase.
\end{abstract}

\maketitle

\section{\label{sec:level1}Introduction}

Active matter is driven out of equilibrium, dissipating energy at the smallest scale, breaking time-reversal symmetry and detailed balance.  Examples of active matter range from bacteria to birds and animals and artificial Janus colloids and vibrated granular material as well. The last few decades have seen tremendous progress in understanding their collective properties and phase transitions~\cite{Marchetti2013, Vicsek2012, Ramaswamy2010, Bechinger2016}.  Theoretical studies of active systems used numerical simulations, kinetic theory, and hydrodynamic theories. In the pioneering Vicsek-model and its variants~\cite{vicsek1995novel, Gregoire2004, Chate2008}, self-propelled particles (SPP) aligning ferromagnetically with the local neighborhood led to flocking transitions. The Toner-Tu theory of coupled orientation and density fields predicted long-ranged order in flocks~\cite{Toner1995a, Marchetti2013}. Extending the equilibrium expectation, microscopic details are often assumed not to affect the emergent macroscopic properties of systems in a given embedding dimension if the symmetries and conservation laws are shared between the models. In contrast, recent studies pointed out changes in macroscopic properties with microscopic implementations in non-equilibrium flocking models with short-ranged interaction~\cite{chepizhko2021revisiting, Pattanayak2018}. For example,  Ref.[\citenum{chepizhko2021revisiting}] showed the dependence on the additivity or the lack of it in the aligning interactions, which is coupled with the presence or absence of reciprocity.   

Note that the Vicsek rule of heading direction aligning with local mean orientation leads to a non-reciprocal torque between a pair of SPPs~\cite{vicsek1995novel}. This rule-based non-reciprocity differs from the emergence of non-reciprocal torques due to the directed motion of polar SPPs, which emerges even when the interactions are modeled as reciprocal~\cite{dadhichi2020nonmutual}. This emergent non-reciprocity can diminish the possible differences between the additive and non-additive flocking models considered in Ref.[\citenum{chepizhko2021revisiting}], as both of them break reciprocity. However, consideration of explicit non-reciprocity can impact other implementations of active systems, e.g., in phase transitions of apolar active nematics~\cite{Simha2002, Ramaswamy2003, Bertin2013a, Mishra2006, chate2006simple, ngo2014large, Das2017a} more profoundly, as the apolar nature of SPPs restricts their directed motion. 
 
 Note that SPPs can lose overall polarity in the presence of fast reversals of self-propulsion direction. In active nematics, a collection of particles align spontaneously along some axis $\hat{\bf n}$ with a $\hat{\bf n} \to -\hat{\bf n}$ symmetry.  Examples of such systems include colliding elongated objects~\cite{peruani2006, ginelli2010large, Peruani2012}, migrating cells~\cite{Gruler1995, gruler1999nematic}, cytoskeletal filaments~\cite{Balasubramaniam2022}, certain direction reversing bacteria~\cite{Wu2009, Theves2013a, Starruss2012, Barbara2003, Taylor1974a}, and vibrated granular rods~\cite{Blair2003, Narayan2007}. 
 In some of these cases, physical processes like an actual collision between active elements may produce alignment~\cite{Blair2003, Narayan2007}, while in others, the effective interaction can be mediated by non-equilibrium means such as more complex biochemical signaling or visual inputs and cognitive processes~\cite{chepizhko2021revisiting, Helbing2000, Saha2020}. Reciprocal interactions can describe former kinds of alignments, while the latter can involve non-reciprocal rules~\cite{Ivlev2015, Fruchart2021, Loos2022}.   
A Boltzmann-Landau-Ginzburg kinetic theory approach~\cite{Bertin2013a} starting from a Vicsek-like implementation of active nematics incorporating non-reciprocal torques led to hydrodynamic equations consistent with earlier results~\cite{Ramaswamy2003, Ramaswamy2010, Marchetti2013}.

In this paper, we explore differences in macroscopic properties of apolar active nematics evolving with short-ranged interaction, particularly their phase transitions, considering three different microscopic implementations: model 1 uses reciprocal torques utilizing a pairwise interaction potential, and models 2 and 3 break reciprocity but in two different manners. The main findings are the following: {  (i)\,the nematic-isotropic (NI) transition is first order (continuous) for reciprocal (non-reciprocal) interaction; (ii)\,the transition from quasi-long-ranged order (QLRO) to disorder is associated with fluctuation dominated phase separation, irrespective of the reciprocity or its absence.}

Section \ref{Models} describes the different apolar nematic models in detail. {  In Section \ref{sec_mft}, we present a mean-field and hydrodynamic analysis complemented by numerical results comparing the NI transitions of the different models. In the following two sections, we analyze the associated phase separation and the nature of the ordered phase.} Finally, we conclude in Section \ref{conclusion} with an outlook.

\section{ Models}\label{Models}
Here, we consider the collective properties of $N$ dry active apolar particles aligning nematically in a 2D area $A=L\times L$. At a given active speed $v_0$, the microstate of particles are described by $\{ \rv_i, \uv_i, q_i \}$, and the particle positions evolve as
\bea 
\rv_i(t+dt) = \rv_i(t)+ q_i\, v_0 \uv_i\, dt.
\label{eq_dr}
\eea
We assume a periodic boundary condition. 
For apolar particles, the polarity $q_i$ is chosen randomly between $\pm 1$ with equal probability. The heading direction $\uv_i = (\cos \h_i, \sin \h_i)$ evolves with the angle $\h_i$ subtended on the $x$-axis. A competition between inter-particle alignment interaction and orientational noise determines the dynamics. For active nematics, alignment interactions are chosen such that the heading direction of neighboring particles gets parallel or anti-parallel to each other with equal probability. In the following, we describe three possible implementations of such alignments that lead to three kinds of macroscopic behavior characterized by differences in the stability of nematic order parameters and particle density.

\subsection{Model 1: Reciprocal model}\label{R-model section}
Within a reciprocal model, we consider that the heading directions of particles interact with the Lebwohl-Lasher potential~\cite{Lebwohl1972, Mishra2006}, $U = -J \sum_{\la ij\ra} \cos [2 (\h_i -\h_j)]$ when inter-particle separations $r_{ij}=|\rv_i - \rv_j|$ remain within a cutoff distance $r_c=1$.  Within the Ito convention, the orientational Brownian motion of $\uv_i(\h_i)$ is described by 
\bea
\h_i(t+dt) = \h_i(t) -\mu (\p U/\p \h_i) dt + \sqrt{2 D_r}\, dB_i(t)
\label{eq_model_1}
\eea
where $\mu$ denotes the mobility, $D_r$ a rotational diffusion constant, and $dB_i$ denotes a Gaussian process with mean zero and correlation $\la dB_i dB_j\ra=\d_{ij} dt$. The equations describe a persistent motion for a free particle where $D_r$ sets the persistence time $\t_p = D_r^{-1}$. In the presence of interaction, the model describes apolar particles aligning nematically with an alignment strength  $J>0$.  Note that the torque felt by a particle pair $i,j$ is equal and opposite to each other.  
{  An earlier on-lattice implementation of this model~\cite{Mishra2006} showed a fluctuation-dominated phase ordering associated with the NI transition.}

Using the Euler-Maruyama scheme, a direct numerical simulation can be performed integrating Eq.(\ref{eq_dr}) and (\ref{eq_model_1}). We use the scaled angular diffusion $D_\h \equiv{2 D_r}/{\mu J}$ as a control parameter to study the properties associated with the NI transition 
setting the number density $\rho={N}/{L^2}=0.4$. The other parameters used in the simulations are $v_0=30$, $dt=0.01$ so that $v_0 dt=0.3$~(see Fig \ref{fig:Svseta}).

\subsection{Model 2: Non-reciprocal model}\label{NR_section}
In Vicsek-like models~\cite{Vicsek2012, chate2006simple}, a particle's heading direction aligns with its neighborhood's mean nematic orientation. The averaging effectively reduces the alignment strength. To capture this behavior approximately within the Hamiltonian scheme, we utilize the mean torque to reorient the heading direction in this model. To this end, we scale the alignment strength by the instantaneous number of neighboring particles $n_i(t)$ within the cutoff radius   
\bea
\h_i(t+dt) = \h_i(t)-(\mu/n_i) (\p U/\p \h_i) dt + \sqrt{2 D_r}\, dB_i(t)
\label{eq_model_2}
\eea
For any $i,j$-pair of particles, the number of neighbors generally does not remain the same, $n_i(t) \neq n_j(t)$. As a result, the interaction strength does not remain reciprocal and breaks additivity. We perform numerical simulations of this model using the same parameters and method as in model 1.  

\subsection{Model 3: Chat\'e model}
Another implementation of non-reciprocal and non-additive interaction, the Chat\'e model for active apolar nematic, was originally proposed in  Ref.[\citenum{chate2006simple}] and discussed further in Ref.[\citenum{chate2020dry}]. Within this model, the heading direction of $i$-th particle aligns with the local average of nematic orientations as   
\bea
\theta_i(t+ dt) &=\Theta (Q_i^t[\{{\rv}(t),\theta (t)\}]) +  \sigma \zeta_i(t)
\label{eq_model_3}
\eea
where $\zeta_i(t)$ is a random number taken from a uniform distribution such as $\zeta_i \in [-\pi/2,\pi/2]$. The implementation of this dynamics is discrete and independent of $dt$ value~\cite{mishra2014aspects}. Using the mean and variance of $\zeta_i$, it is easy to check that the resultant rotational diffusivity of a free mover $D_r=\f{\s^2\pi^2}{24\, dt}$ is proportional to $\s^2$. This expression shows that a continuous time limit of this model with finite $D_r$ and orientation fluctuation $\s$ does not exist. Thus, we present the results of this model in terms of  $\e=\s^2$  to compare with the two other models presented above. 
The operator  $Q_i^t[\{\rv(t),\theta (t)\}])$ determines the local nematic orientation tensor  
\begin{align}
   Q_i^t=& 
   \begin{pmatrix}
  \la\cos 2 \theta_i\ra & \la \sin 2 \theta_i\ra \\
  \la \sin 2 \theta_i\ra & -\la \cos 2 \theta_i\ra
  \end{pmatrix}
\end{align}
where the instantaneous average $\la \dots \ra$ is taken over the instantaneous neighborhood of $\rv_i(t)$ within the cutoff distance $r_c=1.0$. The angle $\Theta$ denotes the orientation of the eigen-direction corresponding to the largest eigenvalue of $Q_i^t$. As noted before, the local averaging around the test particle makes the model non-reciprocal, breaking Newton's third law.  
We perform numerical simulations 
setting density $\rho=0.5$. The active speed is chosen to be $v_0 dt=0.3$, consistent with Ref.[\citenum{chate2006simple}], and the parameter choice in the other two models in this paper. 

\begin{figure}[t]
\centering
  \includegraphics[width=\linewidth]{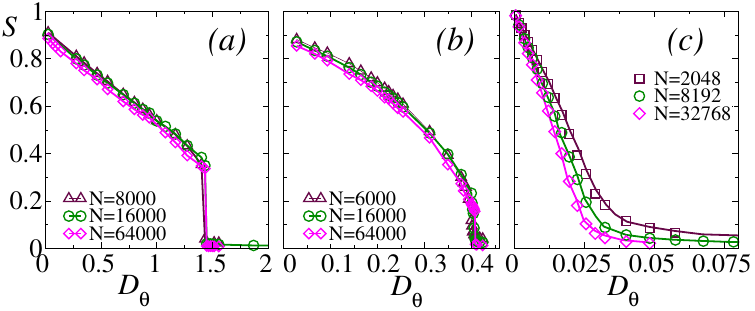}
 \caption{Nematic-isotropic (NI) transition: Variation of the steady-state scalar order parameter $S$ with effective angular diffusivity $\e$ in model 1\, ($a$)  using $\rho=0.4$ and  $N=64000,16000,8000$, model   2\,($b$)  using $\rho=0.4$ and  $N=64000,16000,6000$,  and in model 3\,($c$) using $\rho=0.5$ and  $N=32768,8192,2048$. The lines in 
  ($a$) and  ($b$) are guide to the  eyes while in  ($c$) they stand for earlier results in Ref.[\citenum{chate2006simple}].
 The transition points $\e^{*}=1.43$\,($a$), $\e^{*}=0.407$\,($b$), and $\e^{*} =0.0196$\, ($c$) correspond to maximum standard deviation in $S$.  
  }
  \label{fig:Svseta}
\end{figure}

{ 

From numerical simulations of all three models, we calculate the degree of nematic order using the scalar order parameter, 
\bea
S=[\la\cos(2\theta_i)\ra^2 + \la\sin(2\theta_i)\ra^2]^{1/2} 
\eea
where the average is taken over all the particles $i$ involved. 
For the steady-state average $S$ over the whole system, we take an average over all the $N$ particles and further average over steady-state configurations~(Fig.\ref{fig:Svseta}). For a local calculation of $S$, an averaging over particles in a local volume can be performed. 

\section{Mean field analysis}
\label{sec_mft}
Since orientation dynamics in models 1 and 2 evolve by Langevin equations,  a mean-field analysis of the orientation order using the corresponding  Fokker-Planck approach is straightforward~\cite{chepizhko2021revisiting}. We comment on model 3 at the end of this section. Ignoring density fluctuations, we get 
\bea
&\p_t p (\h,t)= \p_\h\left[ \g (n) \int_0^{2\pi} d\h' \sin[2(\h-\h')] p(\h,t)p(\h',t) \right] \nn\\
&+ D_r \p^2_\h p,
\eea
where $\g (n) = 2 \tilde \mu J n$, with $n = \pi r_c^2 \rho$ denoting the mean number of nearest neighbors. 
It is easy to see from Eq.(\ref{eq_model_1}) and (\ref{eq_model_2}) that $\tilde \mu = \mu$ in model 1 and $\tilde \mu=\mu/n$ in model 2. As a result, 
$\g (n)$ 
increases with $\rho$ in model 1 while  $\g (n) =2J$ remains independent of density in model 2. 
Assuming $\psi$ denotes the direction of broken symmetry and 
\bea
S= \mid \int_0^{2\pi} d\h p_{st}(\h) \exp(i\,2\h) \mid
\label{eq_s}
\eea
denotes the scalar nematic order parameter quantifying the degree of order,  with the steady state distribution of heading directions around the broken symmetry orientation 
\bea
p_{st}(\h) = {\cal N} \exp\left[ \f{\g(n) S}{2 D_r} \cos(2(\h-\psi))\right]. 
\label{eq_p}
\eea
Equations (\ref{eq_s}) and  (\ref{eq_p}) lead to the self-consistency relation $S=I_1\left[ \f{\g(n) S}{2 D_r} \right]/I_0\left[ \f{\g(n) S}{2 D_r} \right]$, with $I_n[.]$ denoting n-th order modified Bessel function of the first kind. For small $S$, performing Taylor expansion, we get $S \approx \f{\g(n) S}{4 D_r}\left(1- \f{\g^2(n) S^2}{64 D_r^2} \right)$. Above the transition point, only one solution exists, $S=0$. Below it,
$S=(2 D_\h/D_\h^{(c)})(1-D_\h/D_\h^{(c)})^{1/2}$ where we used  $D_\h=2D_r/\mu J$
and the critical point 
\begin{equation}
    D_\h^{(c)}=
    \begin{cases}
      n=\pi r_c^2 \rho, & \text{in model 1} \\
      1, & \text{in model 2.}
    \end{cases}
    \label{eq_Dh*}
  \end{equation}
    Thus, within the mean-field analysis, the NI critical point differs in model 1 and model 2. In model 1, the critical point can shift to higher values of scaled angular diffusion with increasing density, a feature absent in model 2.  Similar behavior was predicted before for ferromagnetically aligning polar particles~\cite{chepizhko2021revisiting}.

The above mean-field analysis can not be directly used on model 3, as the orientation evolution in the Vicsek-like model differs from the Langevin description. However, interaction in model 3 is non-additive as in model 2, with spins aligning with the local neighborhood irrespective of the number of neighbors. Thus, the transition is expected to have a density-independent critical point. A smooth variation of $S$ characteristic of the continuous transition is observed in model 3~(Fig.~\ref{fig:Svseta}($c$)\,).   

The discontinuous change in $S$ in  Fig.~\ref{fig:Svseta}($a$) characterizes a first-order transition in model 1. As the figure shows, the size of the discontinuous jump remains unchanged for increasing system size. On the other hand, the discontinuity in $S$ for model 2 decreases with system size $N$, suggesting a weak first-order or continuous transition. 
The two-dimensional NI transition in equilibrium is continuous; however, significant density fluctuations in active nematic can make it first order. In the following, we present a phenomenological hydrodynamic approach~\cite{Ramaswamy2003, Bertin2013a, Das2017a} to explore various possibilities by incorporating the impact of density fluctuation. 
    
  The hydrodynamic theory describes a coupled evolution of slow variables, the particle density $\rho(\rv,t)$ and the local density of nematic order parameter $\Pi_{ij}(\rv,t)=Q_{ij}(\rv,t)\rho(\rv,t)$, with 
  \begin{align}
   Q(\rv,t)=& \f{S}{2} 
   \begin{pmatrix}
  \cos 2 \theta (\rv,t) & \sin 2 \theta (\rv,t) \\
  \sin 2 \theta (\rv,t) & -\cos 2 \theta (\rv,t) 
  \end{pmatrix} \nn
\end{align}
determined by the scalar order parameter $S$ describing the degree of nematic order and orientation $\h(\rv,t)$. 
Using active current $J^{(a)}_i = \l \p_j \Pi_{ij}$~\cite{Marchetti2013}, the particle density field evolves as, 
\bea
\p_t \rho = - \l \p_i \p_j \Pi_{ij} + D \nabla^2 \rho, 
\eea
absorbing the parameter $M$ into $\l$. Here, $D$ denotes the effective diffusivity in the isotropic phase. 
The evolution of nematic order follows
\bea
&\p_t \Pi_{ij} = [\a_1(\rho) - \a_2 {\rm Tr}(\Pi^2) ] \Pi_{ij} - \hf \left[\p_i \p_j - \hf \d_{ij} \nabla^2\right]\rho \nn\\
&+ D_\Pi \nabla^2 \Pi_{ij}.
\label{eq_PI}
\eea
Consistency with the above-mentioned mean-field analysis requires $\a_1(\rho)=D_\h^{(c)}(\rho)-D_\h$,  and $D_\h^{(c)}(\rho)=A\rho$ in model 1 with the area $A$ denoting the range of interaction and $D_\h^{(c)}$ a constant in models 2 and 3.   
A linear stability analysis of the above equations around the homogeneous and isotropic state predicts instability towards the formation of density bands~\cite{Shi2010, Bertin2013a, Das2017a}, a generic feature observed in all our simulations~(Fig.\ref{fig:conf}).

\begin{figure}[t]
\centering
  \includegraphics[width=\linewidth]{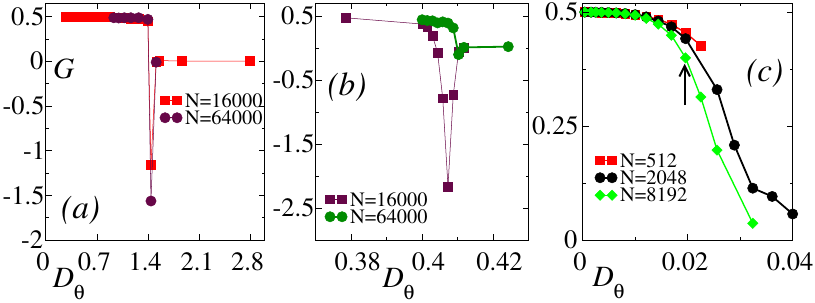}
 \caption{The Binder cumulant $G=1-\la S^4\ra/2 \la S^2\ra^2$: model 1 in ($a$), model 2 in ($b$) and model 3 in ($c$). The arrow in  ($c$) indicates the transition point $D_{\theta}^{(c)}=0.0196$ determined from order parameter fluctuation in Fig.\ref{fig:beta-model-3}($b$).
  }
  \label{fig:G}
\end{figure}

Here, we first analyze the mean-filed NI phase-transition predicted by the above equations, assuming homogeneity with $\rho$ constant and ignoring spatial derivatives in the evolution of $\Pi_{ij}$. This leads to a continuous transition with the scalar order parameter changing from $S=0$ to  
\bea
\rho S= \a_2^{-1/2} [D_\h^{(c)}(\rho)-D_\h]^{1/2}
\eea
in the nematic phase, as $D_\h$ decreases below the critical point $D_\h^{(c)}(\rho)$. The $\rho S$ term in the above expression ensures that the presence or absence of nematic order is subject to the presence of particles in a volume. The mean-field critical exponent is $\beta =1/2$ governing the order parameter $S \sim [D_\h^{(c)}(\rho)-D_\h]^\beta$. 

Now, we proceed to incorporate the impact of density fluctuation in the NI transition~\cite{Das2017a}. 
Note that the first term in Eq.(\ref{eq_PI}) is derivable from a free energy density 
\bea
{\cal F} = -\f{\a_1(\rho)}{2} {\rm Tr} (\Pi^2) + \f{\a_2}{4} {\rm Tr} (\Pi^4), 
\eea
assuming a non-conserved dynamics~\cite{pcmp1995}. 
Consider a small perturbation 
over the homogeneous steady state 
such that $\rho=\rho_0+\d \rho$, with the density fluctuation $\d \rho$ arising due to activity. This leads to $\a_1(\rho) = \a_1(\rho_0)+\a'_1(\rho_0) \d \rho$. When the nematic order itself is small near transition, a zero current steady state condition for the density evolution gives $D \p_j \d\rho = \l \rho_0 \p_i Q_{ij} = \l \p_i \Pi_{ij}$. Thus, the density fluctuation depends on the activity $\l$. Incorporating it into the free energy density, one obtains
\bea
{\cal F}=-\f{u_2}{2} {\rm Tr} (\Pi^2) - \f{u_3}{3} {\rm Tr}(\Pi^3) +\f{u_4}{4} {\rm Tr} (\Pi^4)
\eea
where $u_2=\a_1(\rho_0)$, 
$u_3=3 \l \a'_1(\rho_0)/2D$ and $u_4 = \a_2$. The fluctuation in density generates the cubic term in $\Pi$ in the effective free energy density, resulting in a first-order NI transition.
The first order transition point $D_\h^{\ast}$ is larger than the critical point $D_\h^{(c)}$. 
At this transition point $D_\h^{\ast}
= D_\h^{(c)}
+ \f{(\a_1'(\rho_0))^2}{8 \a_2} \f{\l^2}{D^2}$~\cite{pcmp1995} the scalar order parameter jumps from $S=0$ to 
\bea
S_c=\f{u_3}{6 u_4} = \f{\a_1'(\rho_0)}{4\a_2}\f{\l}{D}.
\eea
This phenomenon is purely active; $S_c$ vanishes linearly with $\l$.  
Between all the three models considered, $\a_1(\rho)$ is density-dependent only in the reciprocal model 1. Thus, the above theory suggests a fluctuation-induced first-order phase transition~\cite{Halperin1974, Chen1978, Binder1987} only in model 1. Fig.~\ref{fig:Svseta} shows a discontinuous change in $S$ in model 1, in agreement with the above prediction. For a small enough system, a similar discontinuity in $S$ is observed in model 2. However, in Model 2, the discontinuity decreases with increasing system size. This suggests a weak first-order to continuous transition for large enough system sizes,  thereby agreeing with the theoretical prediction of continuous transition in model 2. Model 3, on the other hand, shows continuous transition with continuous variation of $S$ across transition for all simulated system sizes.     

\begin{figure}[t]
\centering
 \includegraphics[width=0.9\linewidth]{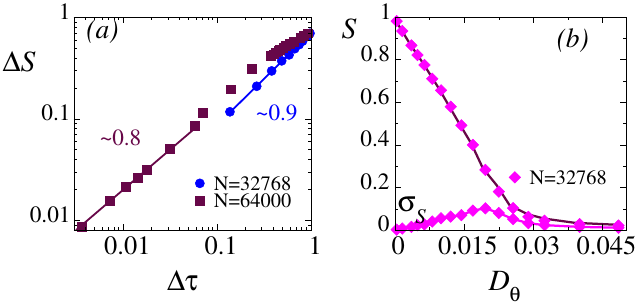}
 \caption{ {To extract the critical exponent, we plot $\D S = S-S^{(c)}$ vs. $\D\tau=1-\e/\e^{(c)}$ below the critical point in model 2 and 3.} 
 $(a)$~$\D S \sim \D \t^\beta$ with $\beta=0.81$ in model 2 (data shown by brown box for $N=64000$) and $\beta=0.92$ in model 3. ($b$)~The critical point $\e^{(c)}$ in model 3 is estimated from the maximum in standard deviation $\s_s = \sqrt{\la S^2\ra - \la S \ra^2}$  in system size $N=32768$. The figure also plots the $\e$-dependence of $S$.  
  }
  \label{fig:beta-model-3}
\end{figure}

{To understand the phase transition further, we compute the Binder cumulant of the scalar nematic order, denoted as $G=1-\la S^4\ra/2 \la S^2\ra^2$, across NI transition for various system sizes~(Fig.\ref{fig:G}). It is defined such that for a Gaussian distribution with mean $\la S \ra=0$, the cumulant vanishes. In a perfectly ordered phase, it approaches 1/2. For a continuous transition, $G$ decreases monotonically with increasing noise, approaching a step function for infinite system size~\cite{Binder1981a}. For a first-order transition, on the other hand, it shows a non-monotonic variation displaying a negative maximum near transition, which gets sharper with system size. This feature reflects the coexistence of ordered and disordered phases in the system~\cite{Vollmayr1993}.}   

Fig.\ref{fig:G} shows clear signatures of first order transition in model 1~(Fig.\ref{fig:G}($a$)\,) and continuous transition in model 3~(Fig.\ref{fig:G}($c$)\,). For larger system sizes, the negative maximum near transition gets sharper and deeper in model 1. In model 3, $G$ for different system sizes merge near the transition point, as below transition, the phase is QLRO~(see  Sec.~\ref{sec_qlro}).  
In model 2, in contrast, a  negative maximum in $G$, characteristic of first-order transition, is observed for smaller system sizes~(Fig.\ref{fig:G}($b$)\,). However, the negative maximum in $G$ gets vanishingly small in larger systems, and the variation of $G$ becomes almost step-function-like as in a continuous transition. This reinforces our earlier conclusion of weak first-order to continuous transition in model 2 for large enough system sizes.
Note that here, we use Binder cumulants to emphasize the difference between the emergent properties of the models, not to find phase transition points.

{
In Fig.\ref{fig:beta-model-3}($a$), we show the numerical estimate of the critical exponent $\beta$ in models 2 and 3 using $\D S = S-S^{(c)} \sim \D \t ^\be$ with $\D\tau=1-\e/\e^{(c)}$ and $S^{(c)}$ the value of $S$ at $\e^{(c)}$. 
In model 3, $\e^{(c)}$ is extracted using the maximum in the variance of $S$~(Fig.\ref{fig:beta-model-3}($b$)\,).  However, in model 2, $\e^{(c)}$ is estimated from the point of step-function like jump in the Binder cumulant observed for the largest system in Fig.\ref{fig:G}.  
We find $\beta \approx 0.81$ for model 2, and  $\beta\approx 0.92$ for model 3. Thus, the two non-reciprocal models show comparable $\beta$ values, which, however, differ considerably from the mean-field estimate of  $\beta=1/2$. A more accurate numerical calculation of $\beta$ requires a careful system-size scaling~\cite{Binder1981a, Binder1981}. Analyzing the coupled hydrodynamics of the density and order parameter fields is necessary for better  theoretical estimation.  
}

 Associated with the NI transition, an instability toward phase separation appears. 
In the following, we establish a fluctuation-dominated phase separation using pair correlation functions and Porod's law violation. Moreover, we show the emergence of QLRO in all three models, using system size scaling of order and calculating nematic correlations.    
}

\begin{figure}[t]
\centering
 \includegraphics[width=\linewidth]{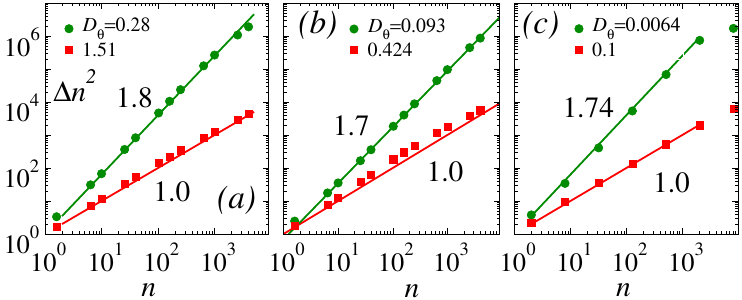}
  \caption{Scaling of the mean-squared number fluctuation  $\Delta n^2 = \la n^2\ra - \la n\ra^2$ corresponding to blocks with mean number of particles $n=\la n \ra$ showing $\Delta n^2 \sim n^{a}$ with $a$ values indicated in the legends. The $\e$ values were chosen such that the value of $S$, in the ordered phase, is comparable for all three models. 
  In ($a$)~model 1: $a=1.8$ for $\e=0.28$, $S=0.78,\,N=64000$, 
  ($b$)~model 2: $a=1.7$ for $\e=0.093$, $S=0.79,\, N=64000$ and 
  ($c$)~model 3: $a=1.74$ for  $\e=0.0064$, $S=0.77,\, N=32768$. 
  Giant number fluctuations~($a>1$) are observed in all three models. The bottom curves in figures ($a$), ($b$), and ($c$) are for the disordered phase in all the models showing normal fluctuations with the exponent $a=1$.  
 }
 \label{fig:gnf}
\end{figure}

\section{Fluctuation dominated phase separation}
\label{phase_sepa}
The coupled dynamics of nematic orientation and local density evolve together. 
The prediction of giant density fluctuation in ordered phase~\cite{Simha2002, Ramaswamy2003} was verified earlier in numerical simulations~\cite{Mishra2006, chate2006simple} and is also observed in the current models (Fig.\ref{fig:gnf}). 
{  We calculate the fluctuation of the number of particles $n$ in a subsystem, $\Delta n^2$, to understand the nature of density fluctuations in more detail. We observe   "giant number fluctuations,"  i.e., $\Delta n^2 \sim n^a,\, a>1$~\cite{Ramaswamy2003,chate2006simple,Marchetti2013} in all three models (Fig.\ref{fig:gnf}) in the ordered phase and normal scaling of fluctuations, i.e., $a=1$ the disordered phase.} 

\begin{figure}[t]
\centering
  \includegraphics[width=8.6cm]{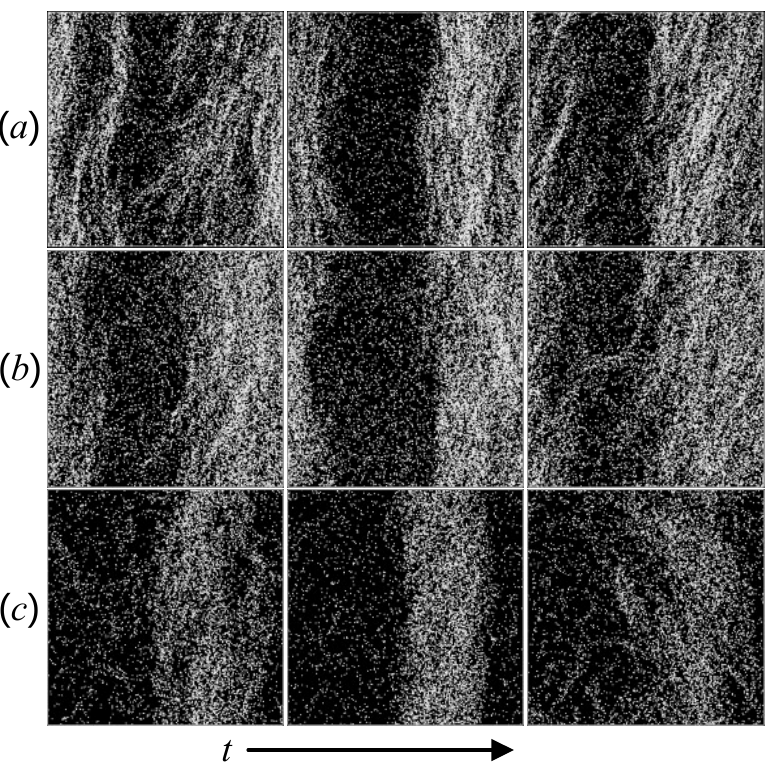}
    \caption{Typical configurations in the density coexistence region. In the grey-scale plot, dark (light) regions denote low (high) density. In panel ($a$)~Model 1: configurations at $\e=0.28,\, \rho=0.4,\, N=16000$,  panel ($b$)~Model 2: configurations at $\e=0.187$ and $\rho=0.4,\, N=16000$, panel ($c$)~Model 3: configurations at $\e=0.012,\,\rho=0.5,\,N=8192$. Fluctuation-dominated bands are observed in all three models.}
 \label{fig:conf}
 \end{figure}

The ordering transition proceeds via the formation of high-density nematic bands. In Fig.\ref{fig:conf}, we show typical configurations inside the nematic phase corresponding to the three models.    
These bands are dynamic; they form and disappear. The interfaces of high and low-density regions show strong fluctuations. All three models show similar behavior. 

\begin{figure}[h!]
\centering
  \includegraphics[width=\linewidth]{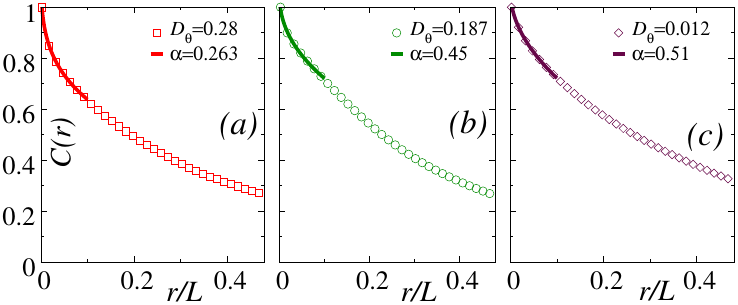}
  \caption{Pair correlation functions  $C(r)=g(r)/g(0)$ for  $\e$  values corresponding to coexistence showing cusps near $r/L \to 0$, indicating a violation of Porod's Law.  The lines show fitting to $C(r)=1-b(r/L)^\a$. Other parameters used are: ($a$) and ($b$) $N=16000,\, L=200$, ($c$) $N=8192,\, L=128$.}
  \label{fig:gr}
\end{figure}

{  To quantify the nature of phase separation more precisely, we calculate the scaled pair correlation function $C(r)=g(r)/g(0)$. 
We observe a cusp at $r/L \to 0$ with $C(r)=1-b (r/L)^\a$ with $\a<1$ (Fig.\ref{fig:gr}) indicating a violation of Porod's law~\cite{Mishra2006, Das2000}, in all three models. The roughness of interfaces of the bands is highest in model 1 among the three models, as evidenced by the largest value of ($1-\alpha$) for this model (see Fig.\ref{fig:gr}), while model 2 and 3 have comparable values of $\a$.}

\section{Quasi long-ranged order}
\label{sec_qlro}
\begin{figure}[t]
\centering
  \includegraphics[width=\linewidth]{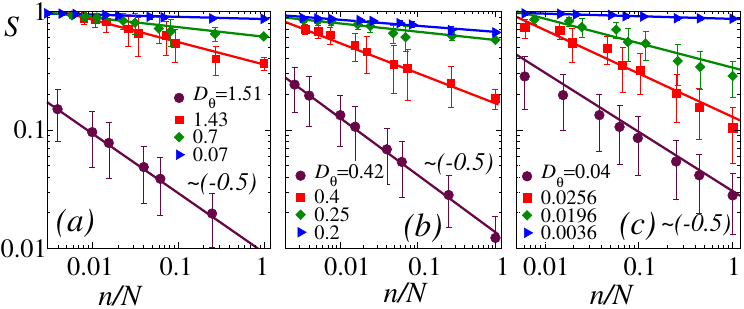}
  \caption{Nematic order parameter $S$ as a function of particle number $n/N$ where $n$ is the average number of particles in a subsystem and $N$ is the total number of particles, at different $\e$ indicated in the legends. ($a$)~model 1:  $N=64000,\rho=0.4$, ($b$)~model 2: $N=64000,\rho=0.4$,  ($c$)~model 3: $N=32768,\rho=0.5$. The solid lines are fitted power laws. As noise $\e$ increases, the power law exponents decrease towards $(-0.5)$ characterizing the disordered phase. 
   }
  \label{fig:SvsN}
\end{figure}
{  To analyze the nature of order, we first calculate the scaling of nematic order $S$ with the number of particles $n$ in subsystems. This shows a power law decay of nematic order $S \sim n^{-\gamma(\e)}$\cite{chate2006simple}. The decay exponent $\g$ increases with $\e$~(Fig.\ref{fig:SvsN}). Deep inside the ordered phase, $\g$ can be vanishingly small as $\e\to 0$. For $\e>\e^{(c)}$, $\g=1/2$ characterizing the completely disordered phase. These features are observed in all three models.}

\begin{figure}[t]
\centering
    \includegraphics[width=\linewidth]{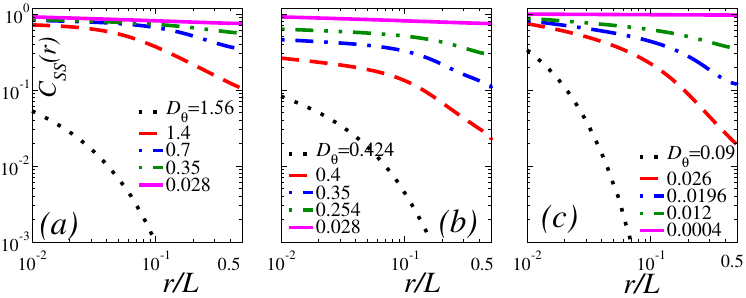}
 \caption{Nematic correlation function $C_{SS}(r)$ plotted at various $\e$ values mentioned in the figure legends. ($a$)~model 1,($b$)~model 2, ($c$)~model 3 show power-law decay of  $C_{SS}(r)$ in the ordered phase, which changes to exponential decay in the disordered phase. Near transition, a cross-over from a short-ranged algebraic to longer-ranged exponential decay is observed. Other parameters are $N=16000,\, L=200$ in ($a$), ($b$) and $N=8192,\, L=128$ in ($c$).  
 }
  \label{fig:Css}
\end{figure}

To further investigate the nature of the order at different phases, we compute the spatial correlation of nematic order~(Fig. \ref{fig:Css}) 
\bea
C_{SS}(r)=\left\la \f{ \sum_{j,k} \cos[2(\theta_j -\theta_k)] \delta(r-r_{jk})}{\sum_{j,k} \delta(r-r_{jk})} \right\ra
\eea
where $r_{jk}$ denotes the separation between a particle pair $j,k$. 
In the ordered phase $C_{SS}(r)\sim r^{-\nu}$, a power law decay characterizing QLRO. 
The decay exponent $\nu$ increases with increasing $\e$, similar to the increase in this exponent with reducing elastic constant in the QLRO phase of XY-spins undergoing Kosterlitz-Thouless transitions~\cite{pcmp1995}. In the disordered isotropic phase at large $\e$,  $C_{SS}(r)$ shows exponential decay, again similar to the disordered phase in Kosterlitz-Thouless systems. However, at intermediate $\e$ values, near the phase transition, $C_{SS}(r)$ shows a behavior unlike the  Kosterlitz-Thouless scenario -- it crosses over from short-ranged algebraic decay to a long-ranged exponential decay. 
The crossover can be explained by noting the presence of high-density bands, although fluctuation-dominated, coexisting with low-density disordered phase. The nematic correlation decays slowly inside the bands and crosses over to exponential decay as one crosses the interface to the low-density regions outside the bands. The above features are common to all three models~(Fig. \ref{fig:Css}).

 Thus,  from the perspective of order-parameter correlation and scaling of nematic order, we recover a semblance of universality in active apolar nematics, with the nature of order in different phases independent of the microscopic implementations.

 \section{Conclusions}\label{conclusion}
  We have shown how differences in microscopic implementations in models that retain the same symmetry lead to qualitatively distinguishable features of phase transitions in active matter. For this purpose, we used three different models for apolar active nematics evolving with short-ranged interaction. Among them, model 1 uses reciprocal torques between active agents. On the other hand, models 2 and 3 have a shared property; both break the microscopic reciprocity, implementing the nematic alignment rules in two different manners. In model 2, an active agent encounters a mean torque due to its neighborhood. On the other hand, in model 3, an active agent aligns with the local mean nematic orientation. Although the models share the same symmetry, as it turns out, {  depending on the presence or absence of reciprocity, they show qualitatively distinguishable phase transitions.} 

  {  
  We employed a mean-field analysis and a phenomenological hydrodynamic theory to demonstrate that reciprocal interactions result in a critical point that gets influenced by density fluctuations while non-reciprocal interactions lead to a critical point oblivious of such fluctuations. Consequently, fluctuations in density can induce a first-order transition in model 1, while models 2 and 3 are expected to exhibit continuous ordering transitions. Our numerical simulations provide compelling evidence supporting this prediction.}

 In Model 1, which incorporates a reciprocal Lebwohl-Lasher interaction~\cite{Lebwohl1972}, a first-order transition is observed between the nematic and isotropic phases. This transition is characterized by {  a sudden, discontinuous change in the order parameter and a characteristic negative peak in the Binder cumulant near the transition point. The microscopic reciprocity is broken in Models 2 and 3. Model 2 shows signatures of weak first-order to continuous transition as the system size increases. In contrast, Model 3 demonstrates a continuous transition across all considered system sizes.  
 
 Despite the above qualitative difference in the ordering transition, all three models show similar properties in the following sense. The ordering transition in all of them proceeds with fluctuation-dominated phase separation. Moreover, the nematic phase shows a quasi-long ranged order in all three models.
 
 The criticality in models 2, 3 are characterized by $S\sim (D_\h^{(c)}-D_\h)^\be$ for $\e < \e^{(c)}$ with the mean-field prediction of the critical exponent $\be=1/2$. The approximate numerical estimates of $\beta$ in the two non-reciprocal models show similar values, which, however, remain considerably larger than the mean-field estimate. To obtain a more accurate theoretical prediction, it might be necessary to examine the full coupled  hydrodynamics of both the density and order parameter fields, and a system size scaling in numerical simulations.} 
 
In conclusion, we have shown that models sharing the same nematic symmetry but with reciprocal and non-reciprocal alignment interactions lead to distinct macroscopic features. The reciprocal model shows a first-order NI transition, while the non-reciprocal models undergo a continuous NI transition. Thus, our findings raise questions on the prevailing notions that macroscopic properties like phase transition should be independent of particular microscopic realizations if they share the same dimensionality, symmetries, and conservation laws.  
Here,  a caveat is in order: it is possible that changing the microscopic models lets one move significantly through a phase space that allows all the different kinds of phase transitions described above.  This possibility cannot be entirely excluded without further studies. 

\section*{Author Contributions}
 DC designed the study. AS performed all the calculations under the supervision of DC. DC wrote the paper with help from AS.

\section*{Conflicts of interest}
 There are no conflicts to declare.

\section*{Acknowledgments}
 D.C. thanks Sriram Ramaswamy and Shraddha Mishra for useful discussions,  SERB, India, for financial support through grant number MTR/2019/000750, and International Center for Theoretical Sciences (ICTS-TIFR), Bangalore, for an Associateship. SAMKHYA, the High-Performance Computing Facility provided by the Institute of Physics, Bhubaneswar, partly supported the numerical simulations.




\bibliographystyle{rsc}

 
\providecommand*{\mcitethebibliography}{\thebibliography}
\csname @ifundefined\endcsname{endmcitethebibliography}
{\let\endmcitethebibliography\endthebibliography}{}

\end{document}